\documentclass[review]{elsarticle}

\pdfoutput=1
\usepackage[latin1]{inputenc}
\usepackage{amsmath}
\usepackage{amsfonts}
\usepackage{amssymb}
\usepackage{graphicx}
\usepackage{subcaption}
\usepackage{natbib}
\usepackage{color}
\usepackage[left=2.50cm, right=2.50cm, top=2.50cm, bottom=2.50cm]{geometry}

\biboptions{sort&compress}

\graphicspath{{./}{./}}

\newcommand{\diffyx}[2]{\frac{{\rm d}#1}{{\rm d}#2}}

\begin{document}

\begin{frontmatter}

\title{A viscous froth model adapted to wet foams}

\author[aber,ums]{Denny Vitasari}
\ead{dsv@aber.ac.uk/denny.vitasari@ums.ac.id}
\author[aber]{Simon Cox\corref{cor1}}
\ead{sxc@aber.ac.uk}
\cortext[cor1]{Corresponding author}
\address[aber]{Institute of Mathematics, Physics and Computer Science,\\
Aberystwyth University, UK}
\address[ums]{Department of Chemical Engineering, \\ 
Universitas Muhammadiyah Surakarta, Indonesia}

\begin{abstract}

We describe the extension of a ``viscous froth'' model to the dynamics of a wet foam in a Hele-Shaw cell. The two-dimensional model includes the friction experienced by the soap films as they are dragged along the cell walls, while retaining accurate geometrical information. To explore the consequences of changing the liquid content in this situation, we consider a simple foam geometry known as a bubble lens: a bubble partially filling a narrow, straight channel with a single film spanning the gap between the bubble and the opposite wall. The triple vertices of this structure are decorated with Plateau borders whose area determines the liquid fraction of the foam.

We derive new expressions to allow the pressure in the Plateau borders to be calculated, and determine numerically the range of driving velocities for which the system reaches a steady state. As the liquid fraction increases, the lens moves more slowly and the spanning film is more greatly distorted, reducing the range of stable driving velocities. For higher velocities, the spanning film moves so quickly that it leaves the bubble behind, a situation which must be avoided in any particular application. 

\end{abstract}

\end{frontmatter}

\section{Introduction}

Aqueous foam consists of air bubbles separated by thin liquid films \cite{Weaire1999, Cantat2013}. Foam has wide industrial applications, ranging from flotation in the mining industry, oil recovery, and foam fractionation, as well as consumer products \cite{Stevenson2012}. Among those applications, aqueous foams flowing through microfluidic channels have received attention during the last decade \cite{Garstecki2004}, supported by advancing technologies in the field \cite{Drenckhan2005}.  Applications of liquid foams in microfluidic channels include  medical, pharmaceutical, and biological fields \cite{Drenckhan2005} as well as oil recovery and soil remediation \cite{Jones2013}. Control of these systems is largely driven by the geometry and flow-rate of the foam and the geometry of the channel, and so mathematical modelling of the situation must move beyond the quasi-static models of the dynamics of dry foams that have been used in the past, to capture how the deformation of the foam depends on its flow-rate and liquid content.

Determining the foam dynamics in a fully three-dimensional model can be very complex and computationally expensive \cite{Green2006, Reinelt1996, Reinelt2000, Reinelt1993,Edwards1991}. Therefore, to reduce the complexity of the system, models of the dissipative dynamics of flowing foams are often applied to two-dimensional foams, i.e. a monolayer of bubbles between two parallel glass plates, both theoretically, for either ordered \cite{Khan1986,Kraynik1987} or disordered \cite{Cox2015, Cox2005, Drenckhan2005, Kern2004, Green2006, Grassia2008,Cantat2011} foams, and experimentally \cite{Geraud2016, Jones2013, Dollet2014, Cantat2014, Marmottant2009}. 

The viscous froth model of Kern et al. \cite{Kern2004} describes the flow of a two-dimensional foam in situations where the dominant source of dissipation is due to friction with the glass plates themselves. In this model the foam is assumed to be in the ``dry limit'' of zero liquid fraction, and the liquid films separating the gas bubbles are very thin. The foam therefore consists of two elements: thin films and the vertices where the films meet in threes \cite{Weaire1999}. The model applies to fast-moving foam \cite{Embley2011b}, where the rate of film deformation is faster than that of mechanical relaxation. Consequently the foam is not in mechanical equilibrium, and the films experience continuous deformation \cite{Green2009}.  

The viscous froth model integrates three physical phenomena~\cite{Kern2004}: 
\begin{itemize}
 \item the viscous drag force resulting from moving the soap films along the confining plates. 
 \item the pressure difference $\Delta p$ across the films. 
 \item the surface tension $\gamma_f$ acting along the films. 
\end{itemize}
The model then describes the balance of forces acting locally at each point on a film, resolved in the direction of the normal. If the film has curvature $C$ and separates bubbles $ b $ and $ b' $ then 
\begin{equation}
\Delta P_{b b'} - \gamma_f C_{b b'} = \lambda v_n,
\label{eq:vf}
\end{equation}
where $ \lambda $ is the viscous drag coefficient and $v_n$ is the velocity in the direction normal to the interface. 

Previously, several authors have used the viscous froth model to examine the case of dry foams \cite{Kern2004,Cox2005,Grassia2008,Green2006,Cox2009}, in which the size of the liquid-carrying Plateau borders, where three films meet, is negligible. This idealisation of real foams will break down as soon as the liquid content reaches a few percent of the total volume \cite{Bolton1991,Weaire1999b}. Here we extend this viscous froth model to wet foams, of arbitrary liquid fraction, and ascertain the effect of liquid on the movement of the foam.

\section{Numerical Method}

\subsection{Wet two-dimensional foams}

\begin{figure}
\begin{subfigure}{.66\textwidth}
  \includegraphics[width=\linewidth]{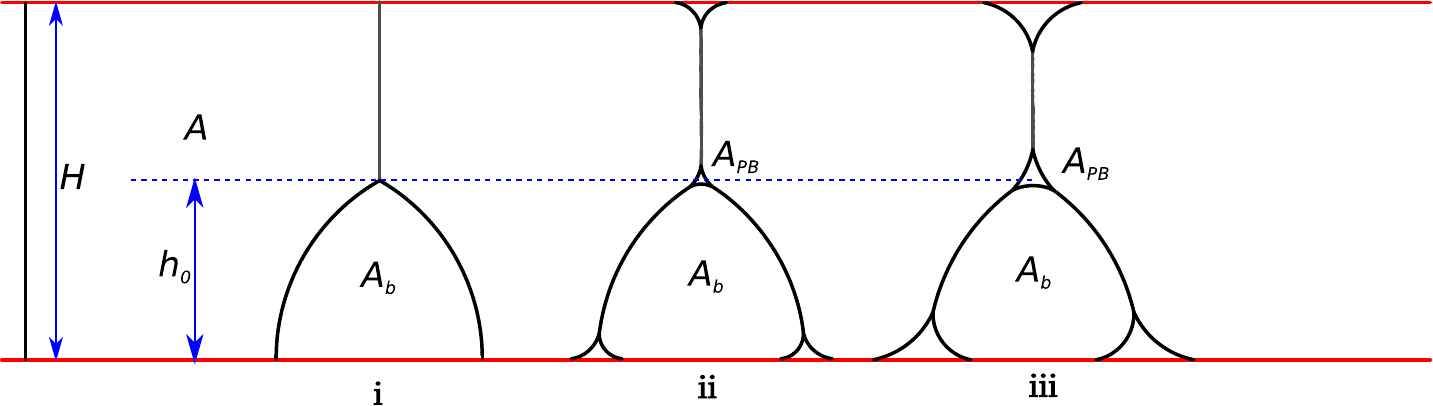}
  \caption{}
\end{subfigure}
\begin{subfigure}{.32\textwidth}
  \includegraphics[width=.973\linewidth]{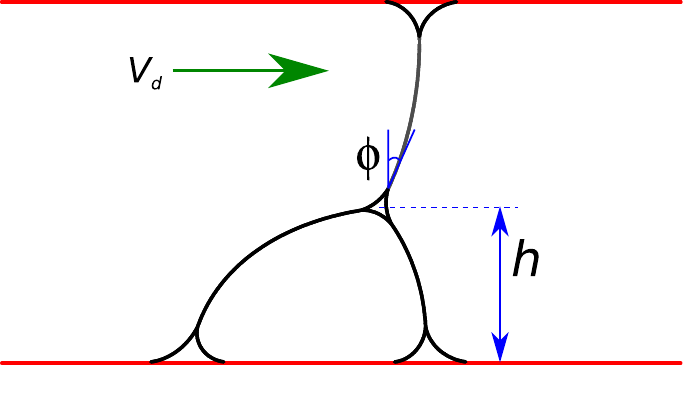}
  \caption{}
\end{subfigure}
\caption{(a) The static structure of a lens bubble. We show three different cases: (i) dry case, without Plateau borders; (ii) and (iii) wet case, in which the lens is decorated with Plateau borders -- one bulk Plateau border and three wall Plateau borders -- with two different Plateau border sizes relative to the lens size: (ii) $ A_{PB}/A_b = 0.005 $ and (iii) $ A_{PB}/A_b = 0.01 $. (b) The shape of the bubble after evolution with the viscous froth model to a steady-state shape. The parameters of this particular simulation are: lens area $ A_b = 0.205 $, driving velocity $ V_d = 2 $, relative size of Plateau border $ A_{PB}/A_b = 0.01 $, contact angle $ \theta = 0.18 $ radians. The angle through which the spanning film has turned is denoted $\phi$ and the linear height of the bubbles is denoted $h$.}
\label{fig:bubble}
\end{figure}

To create a wet foam we replace, as shown in Fig.~\ref{fig:bubble}(a), each triple junction of films with a triangular liquid region to represent each Plateau border. There are two cases to consider. Away from the side walls, there are bulk Plateau borders with approximate three-fold symmetry. On the side walls, the wall Plateau borders meet the wall with a contact angle close to zero. We assume that there is a liquid network that connects all the Plateau borders to a reservoir, so that they all have the same liquid pressure, which is lower than within the bubbles that make up the foam. In contrast to the double interfaces that form the soap films separating two bubbles, the Plateau borders are bordered by single interfaces of tension $\gamma_{PB}$. The sides of the Plateau borders are therefore curved and two of them meet at a vertex to balance the tension in a foam film.

The area of a bulk Plateau border with radius of curvature $r_{PB}$ is $A_{PB} = \left(\sqrt{3} - \frac{\pi}{2} \right) r_{PB}^2$; a wall Plateau border has area $A_{WPB} = \frac{1}{2}(4-\pi)r_{PB}^2$. Since the radius of curvature is linked, {\it via} the Laplace-Young Law, to the pressures, the wall Plateau borders have more than twice the area of the bulk ones.

We choose a particularly simple system in which to study the effect of finite liquid content on the dynamics predicted by the viscous froth model. It consists of just one bubble, a ``lens'' adjacent to one side wall, connected to the opposite wall with a single film -- see Fig.~\ref{fig:bubble}(a). Part of the attraction of this system lies in the extensive calculations \cite{Grassia2008} that exist in the dry case, allowing us to validate our model in the dry limit, where the Plateau borders shrink to zero area.

Following \cite{Grassia2008}, we push the foam along a straight channel of width $H$. To do so we create one further bubble that spans the whole channel; this driving bubble is fixed at one end (denoted by a vertical line in Fig.~\ref{fig:bubble}(a)) and we increase its area $A$ at a fixed rate. The lens is then driven along the channel with an average speed $V_d = \displaystyle \frac{1}{H} \diffyx{A}{t}$.

The computation is carried out using the Surface Evolver software \cite{Brakke1992}. Each interface is discretized into short segments, with lengths in the range 0.005 to 0.015. The curvature of any ``point'' separating two segments is determined by the angle deficit (in practice we use the difference in the direction in which the surface tension force acts on adjacent segments) in going from one segment to the other, normalised by the average length of the two segments. 

An equilibrium structure with the given bubble and Plateau border areas is first found using the Evolver's in-built length-minimization methods. These are then turned off and the area of the driving bubble is slowly increased, with Eq.~\ref{eq:vf} used as an evolution equation for each point (with position $\underline{x}$) by writing the velocity as $\underline{v} = \displaystyle\diffyx{\underline{x}}{t}$, and resolving the result in the direction of the normal to the film. We fix $H=1$ and the time-step is $dt = 1\times 10^{-5}$.

The calculation is carried out in dimensionless form using 
\begin{equation}
\Delta \tilde{P} - \tilde{\gamma} \tilde{C} = \tilde{v}_n,
\label{eq: viscous-froth-dimensionless}
\end{equation}
where $ \tilde{P} = P H / \gamma_f $ is the dimensionless pressure, $ \tilde{C} = C H $ is the dimensionless film curvature, and $ \tilde{v}_n = v_n H \lambda / \gamma_f $ is the dimensionless normal velocity of the film. $ \tilde{\gamma} = \gamma / \gamma_f $ is the dimensionless surface tension, in which $\gamma$ represents either the film or the Plateau border interfaces; thus $ \tilde{\gamma} $ takes the value one for films and just over one half (details are given below) on the walls of the Plateau borders. Henceforth we use the dimensionless versions of the variables and drop the $\tilde{\;\;}$; values of the main independent variables varied in the simulations are given in Tab.~\ref{tab: variables}.

\begin{table}[]
\centering
\caption{The main independent variables varied in the simulations}
\label{tab: variables}
\begin{tabular}{lll}
\hline 
Variable          & Symbol                                         & Range of values \\
\hline 
Contact angle                   & $ \theta $                          & $0.11 - 0.43$ radians \\
Lens bubble area        & $ A_b $                                    & 0.074 - 0.295 $H^2$  \\
Driving velocity        & $V_d =\displaystyle \frac{1}{H} \frac{\mathrm{d}A}{\mathrm{d}t}$      & 1 - 10 \\
Ratio of Plateau border to bubble area        & $A_{PB}/A_b$      & 0.005 - 0.02 \\
\hline
\end{tabular}
\end{table}

We now describe the calculation of the pressure difference between bubbles ($ \Delta P $), required in Eq.~\ref{eq:vf}, extend it to the calculation of the pressure in the Plateau borders, and then give the rule for moving the vertices that connect three interfaces.

\subsection{Determination of the pressures}
\label{sec: pressure}

To calculate the pressure difference between bubbles and between bubbles and Plateau borders, we first calculate the pressure within each bubble and each Plateau border, up to an additive constant. The principle of the method is that any change in bubble pressure should lead to corresponding changes in the bubble area. As in \cite{Kern2004}, we integrate the equation of motion (Eq.~\ref{eq:vf}) around the perimeter $ \partial b $ of a bubble $ b $, now extending this idea to the perimeter of each Plateau border too.

We note first that the change of the area $A$ of a region is equal to the integral of the normal velocity $v_n$ around it, and hence: 
\begin{equation}
\lambda \frac{\mathrm{d}A}{\mathrm{d}t} = \lambda \oint_{\partial b} v_n \mathrm{d}l =   \oint_{\partial b} (\Delta P - \gamma C) \mathrm{d}l,
\label{eq: pressure-1}
\end{equation}
where we denote by $\mathrm{d}l$ an element of length of the interface separating two regions. 

The integral of the curvature can be calculated in terms of the deficit in the turning angle $ \theta_i $ at each vertex using Gauss' theorem: 
\begin{equation}
\oint_{\partial b} C = 2 \pi - \sum_{i=1}^V \theta_i,
\label{eq: curvature-1}
\end{equation}
where $ V $ is the number of vertices of the region. 

For a bubble in a dry foam, the turning angles are $ \theta_i = 2\pi/3 $, therefore the deficit is $\pi/3$ and the integral of the curvature is 
\begin{equation}
\oint_{\partial b} C = 2 \pi - V \frac{\pi}{3} = \frac{\pi}{3}(6-V).
\label{eq: curvature-2}
\end{equation}
This equation applies only in the dry limit, and must be adapted for bubbles in which the vertices are decorated with Plateau borders. Indeed, the effect of the Plateau borders is to reduce the turning angles to zero, since the interface runs smoothly from separating bubbles to separating a bubble from a Plateau border. Hence a bubble with Plateau borders at every vertex has $\displaystyle \oint_{\partial b} C = 2 \pi$. Note that the boundary of the driving bubble includes two vertices, with turning angles of $\pi/2$, and three Plateau borders, and we therefore mix these rules accordingly to get $\displaystyle \oint_{\partial b_d} C  = 2 \pi - 2\frac{\pi}{2} = \pi$.

For both bulk and wall Plateau borders, to continue around the boundary at a vertex requires that we turn through an angle of $\pi$. Since each Plateau border has three vertices, the integral of curvature is
\begin{equation}
 \oint_{\partial PB} C  = 2 \pi - 3 \pi = - \pi.
\label{eq: curvature-Pb}
\end{equation}

Each Plateau border interface has tension $\gamma_{PB}$ and we denote its length by $l_{PB}$. Then we can express Eq.~\ref{eq: pressure-1} as
\begin{equation}
\lambda \frac{\mathrm{d}A_{PB}}{\mathrm{d}t} =  - \pi \gamma_{PB} + \sum_{b} (P_{PB} - P_{b}) l_{PB},
\label{eq: pressure-2}
\end{equation}
where the sum is over all bubbles $b$ touching the Plateau border and, where necessary, the downstream gas with zero pressure.

The boundary of a bubble now consists of interfaces with two different surface tensions. To make the calculation of the bubble pressure tractable, we make the approximation in Eq. \ref{eq: pressure-1}, on the basis that the Plateau borders are small compared to the bubbles, that the tension of the interfaces making up the boundary of a bubble are all equal to $\gamma$. Then, as in Eq. \ref{eq: pressure-2}, we obtain from Eq. \ref{eq: pressure-1} an equation relating the pressures to known geometric quantities $\displaystyle \oint_{\partial b} C$ and the interface lengths. 

We set the area change to be zero for all regions except the driving bubble, which has a fixed increase in area as described above, and solve this matrix equation to give all pressures in the system.

\subsection{Movement of three-fold vertices}

In the dry case \cite{Kern2004}, the three-fold vertices where films meet were moved so as to keep $120^\circ$ angles in the bulk, sometimes known as the Fermat-Steiner point \cite{Tong1995, Gueron2002}, and to meet the wall at $90^\circ$. Here, in the wet case, we must instead consider a vertex at which three interfaces of different tension meet, defining the apices of the Plateau borders. We consider the two cases, the bulk Plateau borders and those on the wall, separately.

\begin{figure}
\centering
\includegraphics[width=0.9\linewidth]{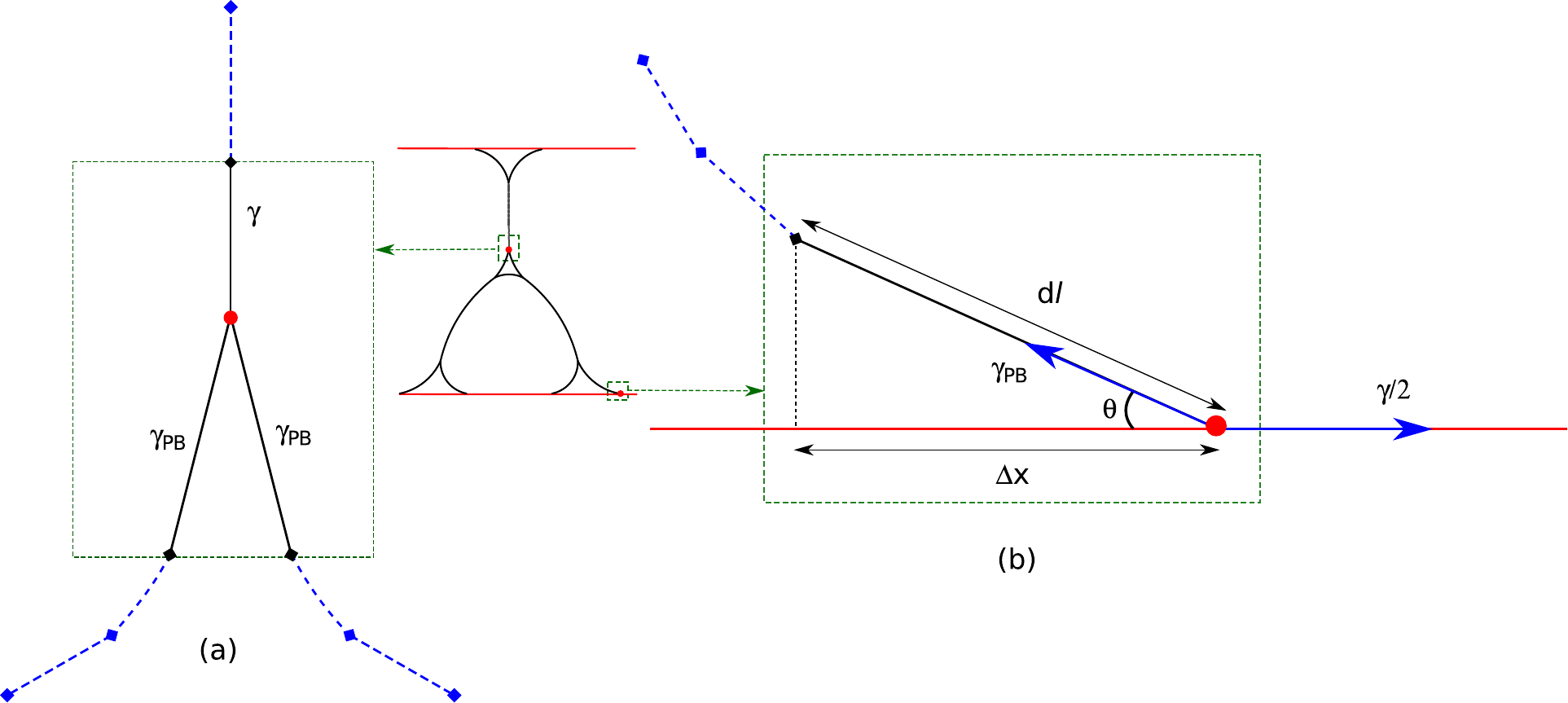}
\caption{Illustration of the vertices, shown with the large red dots) at the corners of the Plateau borders where discretized interfaces meet: 
(a) Solution of the Fermat-Torricelli problem to find the position of a three-fold vertex at one corner of a bulk Plateau border.   
(b) Vertices at the corners of a wall Plateau border are moved in such a way that the contact angle $\theta$ at which the Plateau border interface meets the channel wall is kept constant.}
\label{fig:pb-wall}
\end{figure}

For a bulk Plateau border, we seek instead the Fermat-Torricelli point \cite{Uteshev2014}, with weights given by the three tensions $\gamma, \gamma_{PB}, \gamma_{PB}$. A series of formulae for determining the position of the vertex based on the positions of the three neighbouring points of the discretized interfaces, one in each interface, as illustrated in Fig.~\ref{fig:pb-wall}(a), are given by Uteshev \cite{Uteshev2014}.

For a wall Plateau border we take a different approach, to avoid having to include the otherwise unnecessary film between the Plateau border and the wall, which would slow down the simulation. The contact angle $\theta$ between the Plateau border and the channel wall, shown in Fig.~\ref{fig:pb-wall}(b), is a result of the balance between the surface tension of the wetting film on the wall and the Plateau border tension $\gamma_{PB}$. The tension on the wall is half the tension of the film since at the wall the film has only one surface. That is, $\gamma/2 =  \gamma_{PB} \cos \theta$. Further, the horizontal distance of the position of the vertex from the first point of the discretized Plateau border interface can also be written in terms of $\theta$; since we know the length of the segment connecting them, $\mathrm{d}l$, we move the vertex so that it is a distance $\Delta x = \mathrm{d}l\displaystyle  \frac{\gamma}{ 2\gamma_{PB}}$ to the right or left (depending on which side of the wall Plateau border we are considering) of this point of the discretization. In this way the contact angle remains constant and the vertex remains on the wall.

We choose $\gamma_{PB}$ slightly greater than half of $\gamma$, to give a finite contact angle at all vertices (both wall and bulk).  The contact angle $\theta$ is also half the angle between two Plateau border interfaces where they meet at the vertex of a bulk Plateau border. It is therefore given by
\begin{equation}
\cos \theta =  \frac{\gamma}{2 \gamma_{PB}}.
\end{equation}

\section{Results}

\subsection{Validation in the dry limit}

In the limit in which the area of the Plateau borders shrink to zero volume, we recover the situation considered in \cite{Green2006}. This allows us to validate our simulations, and establish the correspondence between the driving pressure, as used in \cite{Green2006}, and the driving velocity, which we consider here since we know this quantity more accurately. 

\begin{figure}
\begin{subfigure}{.5\textwidth}
  \centering
  \includegraphics[width=\linewidth]{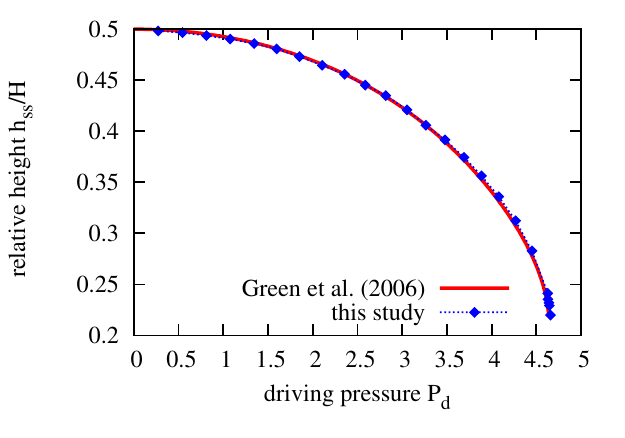}
  \caption{}
\end{subfigure}%
\begin{subfigure}{.5\textwidth}
  \centering
  \includegraphics[width=\linewidth]{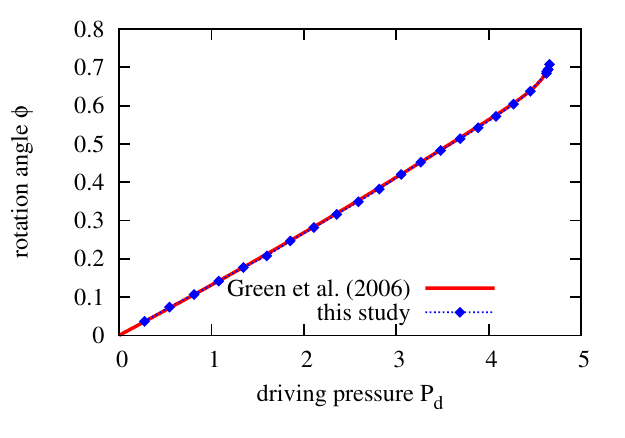}
  \caption{}
\end{subfigure}%

\begin{subfigure}{.5\textwidth}
  \centering
\includegraphics[width=\linewidth]{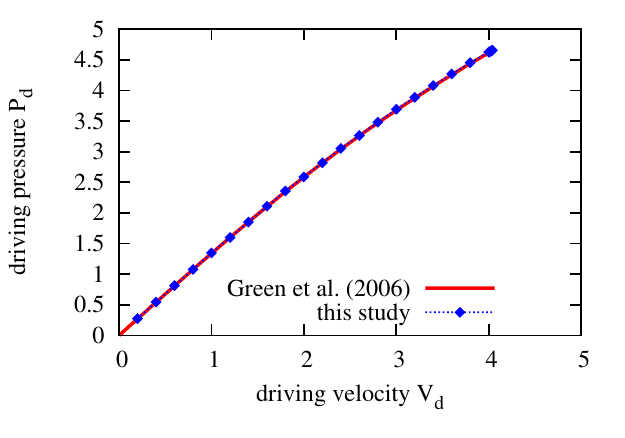}
  \caption{}
\end{subfigure}
\caption{Steady state geometry of the lens bubble in the dry case, compared with the analytical result \cite{Green2006}, for a lens bubble with area $ A_b = 0.205 $. (a) Lens height versus driving pressure. (b) Turning angle versus driving pressure. (c) Correlation between driving pressure (measured) and driving velocity (imposed).}
\label{fig:lens_nopb_grassia}
\end{figure}

The surface tensions of the three films are all the same. We set the initial lens area to be $ A_b = 0.205 $, which corresponds to an initial lens height equal to half the channel width ($ h_0 = 0.5 $), choose a driving velocity in the range $ 0 \leq V_d \leq 5 $ and run the simulation until the vertical position of the triple junction converges to an accuracy of $2\times 10^{-3}$. Then for each value of driving velocity we record both the pressure $P_d$ of the driving bubble and the height of the lens bubble in this steady state. Fig.~\ref{fig:lens_nopb_grassia}(a) shows that the steady state height decreases with increasing driving pressure, as the bubble is more greatly deformed. The data shows excellent agreement with the results of Green et.al \cite{Green2006}. Note that there is a maximum driving pressure, above which no steady state solution is possible: here the bubble deforms sufficiently that the leading edge of the lens bubble shrinks to zero length, and the spanning film detaches from the bubble and proceeds downstream on its own, leaving the bubble behind on the wall. Agreement is also found (see Fig.~\ref{fig:lens_nopb_grassia}(b)) in the turning angle of the spanning film at steady state, up to the critical driving pressure. 

Fig.~\ref{fig:lens_nopb_grassia}(c) shows that the driving pressure and driving velocity are almost linearly related, and it is therefore appropriate to use driving velocity as the independent variable, which we do in the following.

\subsection{Effect of Plateau borders on the bubble shape}

\begin{figure}
\begin{subfigure}{.5\textwidth}
  \centering
  \includegraphics[width=0.9\linewidth]{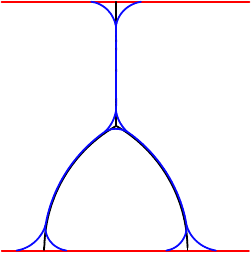}
  \caption{}
  \label{fig: lens_initial}
\end{subfigure}%
\begin{subfigure}{.5\textwidth}
  \centering
  \includegraphics[width=0.9\linewidth]{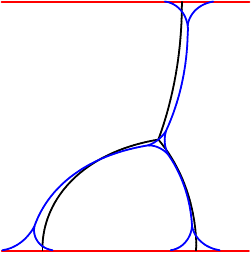}
  \caption{}
  \label{fig: lens_final}
\end{subfigure}
\caption{Bubble shape at different times, with lens bubble area $ A_b = 0.205 $. The dry case is shown with black lines. The wet case is shown in blue, with liquid fraction $ A_{PB}/A_b  = 0.01 $ and contact angle $ \theta = 0.18 $ radians. (a) Initial shape. (b) Shape at steady state with driving velocity $V_d = 2 $. The two images are superimposed to  compare the evolution of the lens shape and the rotation angle of the spanning film. }
\label{fig: lens_picture}
\end{figure}

Introducing Plateau borders at the central vertex and at the three vertices where films meet the wall only slightly changes the static shape of the bubble, as shown in Fig.~\ref{fig: lens_picture}(a). This change is because part of the Plateau borders take up space that was formerly occupied by gas. Thus the walls of the lens bubble are pushed outwards slightly, and so the centre of the bulk Plateau border is slightly higher than in the dry case. This effect becomes more noticeable at larger Plateau border area, i.e. at larger liquid fraction.

When the lens bubble is driven along the channel, the presence of the Plateau borders has a more marked effect (Fig.~\ref{fig: lens_picture}(b)). At the side walls of the channel, the previous $90^\circ$ condition is relaxed by the presence of the Plateau borders, allowing the films to emerge from the apex of the Plateau border at a different angle. This means that the films are more curved, and the turning angle of the spanning film increases. The lens bubble also widens and its height decreases compared to the dry bubble at the same driving velocity.

\subsection{Effect of Plateau border area}

\begin{figure}
\centering
\begin{subfigure}{.5\textwidth}
  \centering
  \includegraphics[width=\linewidth]{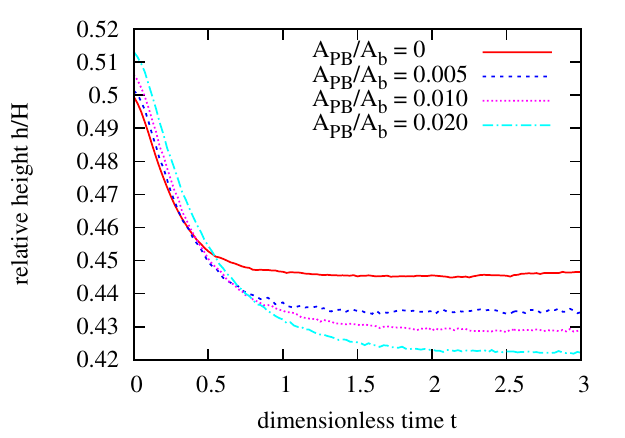}
  \caption{}
\end{subfigure}%
\begin{subfigure}{.5\textwidth}
  \centering
  \includegraphics[width=\linewidth]{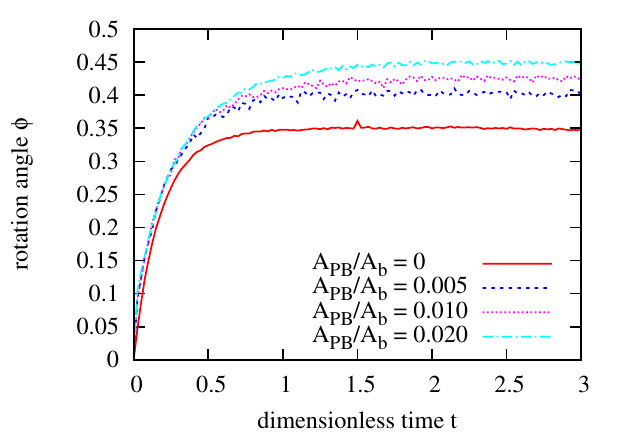}
  \caption{}
\end{subfigure}
\caption{Changes in bubble geometry versus time for various liquid fractions, with lens area $ A_b = 0.205 $, contact angle $ \theta = 0.18 $, and driving velocity $V_d = 2$. The results for different Plateau border areas are compared with the dry case. (a) Height of the lens bubble (position of the centre of the bulk Plateau border). (b) The rotation angle of the spanning film.}
\label{fig:lens_r_y}
\end{figure}

An important parameter in the determination of foam dynamics is the liquid fraction, which is represented here by the area of the Plateau borders relative to the area of the lens bubble. We show in Fig.~\ref{fig:lens_r_y} how two key geometrical parameters of the bubble structure change in time for a given initial height and driving pressure.

In the wet case, we now calculate the height of the top of the lens bubble as the mean vertical position of the three bulk Plateau border vertices. Fig.~\ref{fig:lens_r_y}(a) quantifies the reduction in height of the lens bubble seen in Fig.~\ref{fig: lens_picture}(b). As the lens bubble moves along the channel, its height decreases. Initially, the Plateau border starts in a higher position for given bubble area, and the initial height increases with Plateau border area. But when the bubble is driven along the channel the larger the Plateau border the more the bubble is deformed and the lower its height at steady state.

This is linked to the rotation of the spanning film, which is greater for larger Plateau borders (Fig.~\ref{fig:lens_r_y}(b)). This is because the increase of lens height for larger Plateau borders reduces the length of the spanning film, resulting in greater curvature of that film. The fluctuations in the value of the rotation angle visible in Fig.~\ref{fig:lens_r_y}(b) are a consequence of changes to the discretisation of the spanning film, and in particular the introduction of new points in the lengthening film.

\subsection{Effect of contact angle}

\begin{figure}
\centering
\begin{subfigure}{.5\textwidth}
  \centering
  \includegraphics[width=\linewidth]{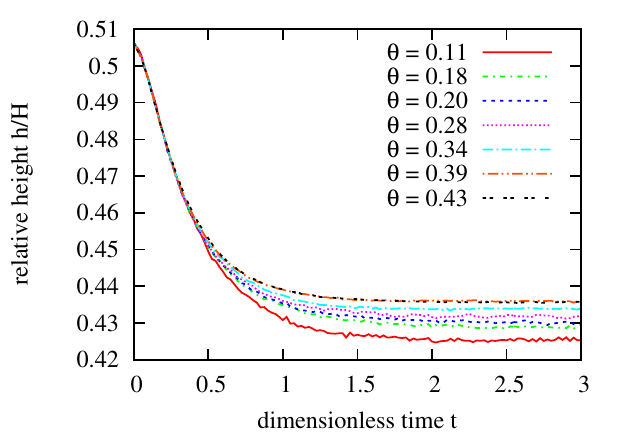}
  \caption{}
\end{subfigure}%
\begin{subfigure}{.5\textwidth}
  \centering
  \includegraphics[width=\linewidth]{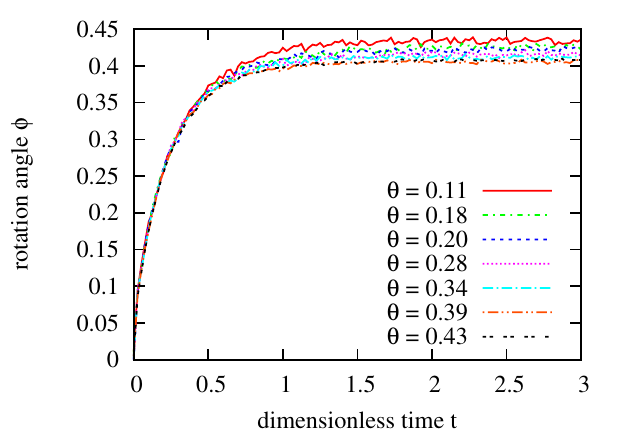}
  \caption{}
\end{subfigure}
\caption{Effect of contact angle on the variation of foam geometry versus time, for a lens bubble with area $ A_b = 0.205 $, liquid fraction $ A_{PB}/ A_b = 0.01 $, driven at velocity $V_d = 2$. (a) Height of the lens bubble (position of the centre of the bulk Plateau border). (b) The rotation angle of the spanning film.  }
\label{fig:lens_gamma}
\end{figure}

In the ideal foam case, the contact angle at the apices of each Plateau border is expected to be zero. As explained above, we relax this condition slightly and introduce a small, finite, contact angle there, by increasing the tension of the Plateau border interfaces. We demonstrate that this contact angle has little effect on the foam geometry here. For contact angles in the range $ 0.11  \leq \theta \leq 0.43 $ radians, Fig.~\ref{fig:lens_gamma} shows that there is little variation in either the lens bubble height or the rotation of the spanning film.  Larger contact angles give slightly higher lens bubbles and reduce the rotation of the spanning film.

\subsection{Effect of lens bubble size}

\begin{figure}
\centering
\begin{subfigure}{.5\textwidth}
  \centering
  \includegraphics[width=\linewidth]{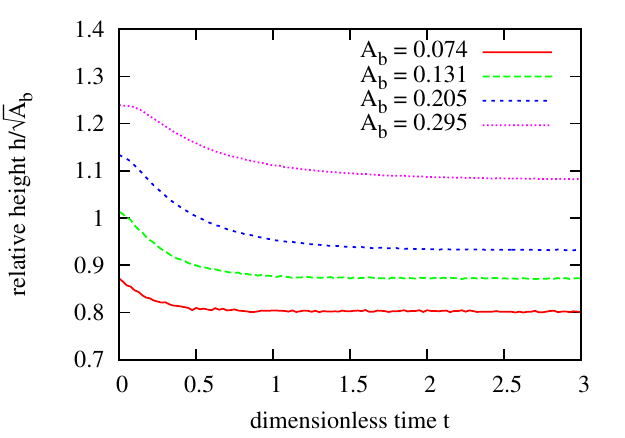}
  \caption{}
\end{subfigure}%
\begin{subfigure}{.5\textwidth}
  \centering
  \includegraphics[width=\linewidth]{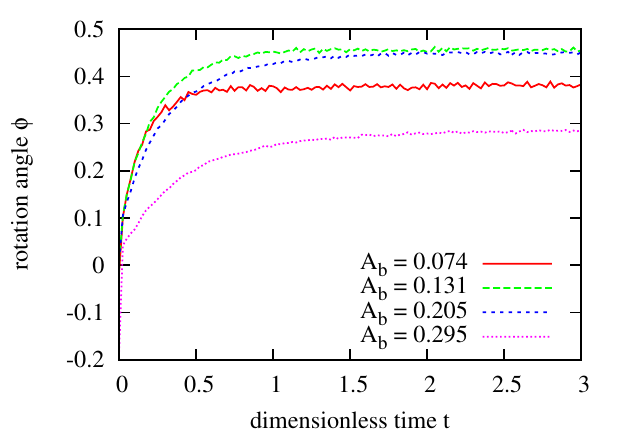}
  \caption{}
\end{subfigure}
\caption{Effect of lens bubble size on the variation of foam geometry versus time, for a lens bubble with contact angle $ \theta = 0.18 $, liquid fraction $ A_{PB}/ A_b = 0.02 $, driven at velocity $V_d = 2$. (a) Height of the lens bubble (position of the centre of the bulk Plateau border) relative to the lens size. (b) The rotation angle of the spanning film.
}
\label{fig:lens_h}
\end{figure}

In our simulations, larger bubbles are surrounded by larger Plateau borders. Nonetheless, by varying the size of the lens bubble we detect a difference in the motion of the foam along the channel. To see this, we normalize the height of the lens bubble by its area, and track the relative height in time, as shown in Fig.~\ref{fig:lens_h}(a). We see that larger bubbles are relatively taller and narrower, while smaller bubbles are shorter and wider. Further, the most rapid decrease in height during the motion is found for the bubbles of intermediate size, which are therefore more deformed by the flow.

Fig.~\ref{fig:lens_h}(b) shows the rotation angle of the spanning film for the same set of simulations, and again we observe that it is the films attached to the bubbles of intermediate size, and therefore spanning films of intermediate lengths, that are most deformed. The greatest angle is found for the bubble with area  $ A_b = 0.131 $, which corresponds to an initial height equivalent to about $h_0 = 0.4$, which is for smaller bubbles than the upper limit found in the dry case.

\subsection{Steady state structure of the foam}

As the graphs above suggest, for many sets of parameters the shape of the bubble reaches a steady state as it moves along the channel. For each lens bubble area we now seek the steady state shape of the lens for each driving velocity up to the critical one. We expect larger lenses to be more greatly deformed by the flow, since they have greater surface area and are therefore affected more by the viscous drag. Changing the lens size changes the length of the spanning film and so, conversely, smaller lenses are attached to longer spanning films, and in this case the shape of the spanning film is more greatly affected by the drag.  Note that we keep the effective liquid fraction fixed, so that larger bubbles are surrounded by larger Plateau borders.

\begin{figure}
\centering
\begin{subfigure}{.5\textwidth}
  \centering
  \includegraphics[width=\linewidth]{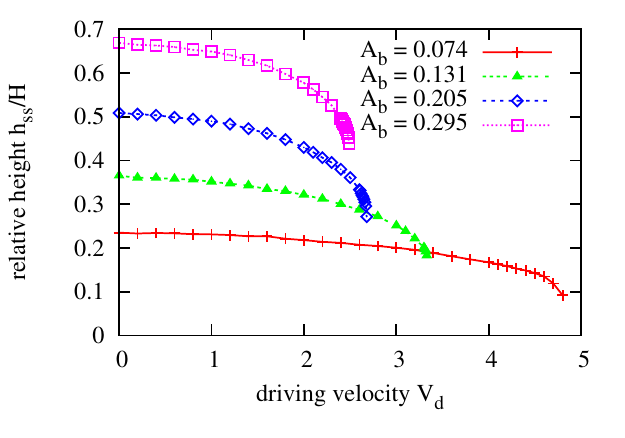}
  \caption{}
\end{subfigure}%
\begin{subfigure}{.5\textwidth}
  \centering
  \includegraphics[width=\linewidth]{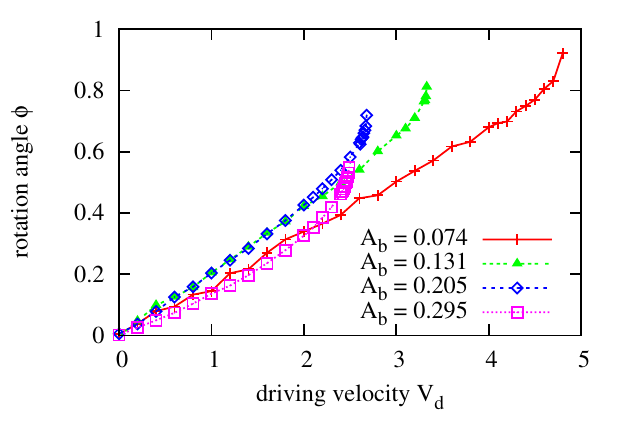}
  \caption{}
\end{subfigure}
\caption{Steady state foam structure for different lens bubble areas $ A_b $ at different driving velocities $V_d$. The contact angle is fixed at $ \theta = 0.18 $ and the liquid fraction is $ A_{PB}/A_b = 0.01 $. Each point on the graphs represents one simulation which runs until the relative height of the lens bubble doesn't change to within $2 \times 10^{-3}$.
(a) Steady state lens height.
(b) Steady state rotation angle of the spanning film. 
}
\label{fig:lens_v3456}
\end{figure}

Fig.~\ref{fig:lens_v3456}(a) shows that the steady state height of larger bubbles reduces more quickly with increasing driving velocity (i.e. the initial slope of the steady state height is more negative for larger lens bubbles). Larger lens bubbles more quickly reach their maximum possible driving velocity, when the two Plateau borders at the leading edge of the bubble meet. Close to this point, the steady state height of the bubble decreases rapidly with increasing driving velocity, as in the dry case \cite{Green2006}.

For small driving velocities, the steady state rotation angle of the spanning film (Fig.~\ref{fig:lens_v3456}(b)) is nearly linear in $V_d$. This rotation angle depends on the lens size and the length of the spanning film: both short and long spanning films deviate less from the vertical \cite{Green2006} than intermediate length films, and hence show smaller rotation angles. At larger driving velocities, there is a limit beyond which the steady solution does not exist, and as this limit is approached the lens height and the rotation angle change significantly with a small change of driving velocity.

\subsection{Critical driving velocity}
\label{sub: max vely}

The critical driving velocity is the steady state velocity at which the bulk Plateau border touches the leading wall Plateau border attached to the lens bubble. Above this driving velocity it is not possible to drive the bubble along the channel; instead, the spanning film will precede it and leave the bubble in a stationary position attached to the wall.

\begin{figure}
\centering
\includegraphics[width=0.7\linewidth]{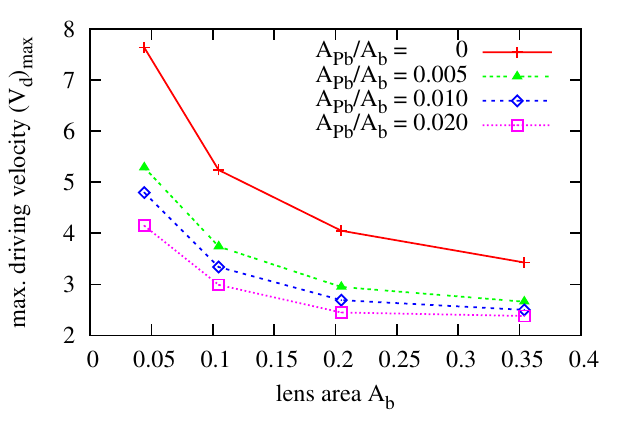}
\caption{The critical driving velocity $ V_d^{\rm max} $, above which the lens can not be driven, decreases with lens size and with liquid fraction. The contact angle is $ \theta = 0.18 $.  }
\label{fig:lens_summary}
\end{figure}

Fig.~\ref{fig:lens_summary} summarises and extends the data from Fig.~\ref{fig:lens_v3456} by recording the critical velocity for various lens bubble sizes and for different liquid fractions. As in Fig.~\ref{fig:lens_v3456}, larger bubbles have lower critical driving velocities \cite{Green2006}. Further, the inclusion of Plateau borders, even small ones with area $ A_{PB}/A_b = 0.005 $ which might correspond to a liquid fraction as low as 0.001 \cite{Gay2011}, significantly reduces the critical driving velocity for all bubble sizes. Increasing the liquid fraction reduces the range of possible driving velocities even more.

\section{Conclusions}

We have extended the viscous froth model of Kern et al. \cite{Kern2004} by adding Plateau borders to allow the simulation of wet foams. The presence of Plateau borders allows the films to curve more, as they relax the constraint of meeting the wall at a right angle. Larger Plateau borders, i.e. higher liquid fraction, result in larger turning angles of the spanning film, as well as greater deformation of the lens bubble when it is driven along the channel at constant velocity. The contact angle at which a Plateau border meets a film has little effect on the deformation of the lens in the range studied here. 

The steady state rotation angle of the spanning film decreases, for given driving velocity, as it becomes shorter or as the lens becomes smaller. In the dry case the maximum rotation angle was found at $ h_0 = 0.5 $, and our results suggest that the presence of liquid in the foam changes this value in such a way that the maximum rotation of the spanning film occurs for smaller lens bubbles (i.e. bubbles with lower initial height).

There is a significant effect of liquid fraction on the velocities at which bubbles can be pushed along channels: the maximum driving velocity drops substantially (at fixed bubble size) meaning that the introduction of a little liquid into a microfluidic system reduces the possible range of bubble sizes that can be used. Thus the friction experienced by a foam with the upper and lower walls of a Hele-Shaw cell has an important effect on the flow.

In this work we have neglected any gradients of surface tension that may arise due to the foam movement. Friction between the Plateau border, the film, and the wetting films on the surfaces of the channel may lead to re-distribution of surfactant, and we plan to include this effect in future work.

\section*{Acknowledgements}

We thank P. Grassia for providing results for the dry case and acknowledge the UK Engineering and Physical Sciences Research Council (EPSRC) for financial support through Grant No. EP/N002326/1.

\section*{References}
\bibliographystyle{elsarticle-num-names}
\bibliography{viscous-froth}

\end{document}